\documentclass[sigconf]{acmart}

\usepackage{tcolorbox}
\usepackage{xcolor}
\usepackage{algorithmic}
\usepackage{graphicx}
\usepackage{enumerate}
\usepackage{subcaption}
\usepackage{caption}
\usepackage{csquotes}
\usepackage[utf8]{inputenc}

\graphicspath{{figs/}}

\definecolor{codegreen}{rgb}{0,0.6,0}
\definecolor{codeblue}{rgb}{0,0,0.6}
\definecolor{codegray}{rgb}{0.5,0.5,0.5}
\definecolor{codepurple}{rgb}{0.58,0,0.82}
\definecolor{backcolour}{rgb}{0.95,0.95,0.92}

\def\isdraft{1}

\newcommand{\mynote}[3]{\fbox{\bfseries\sffamily\scriptsize #1}{\small\textsf{\emph{\color{#3}{ #2}}}}}

\if\isdraft1
\newcommand{\petrillo}[1]{\mynote{Petrillo}{#1}{red}}
\newcommand{\jalves}[1]{\mynote{Jalves}{#1}{codegreen}}
\newcommand{\todo}[1]{\mynote{TODO}{#1}{codepurple}}
\newcommand{\argument}[1]{\mynote{Argument:}{#1}{codeblue}}

\else
\newcommand{\petrillo}[1]{}
\newcommand{\jalves}[1]{}
\newcommand{\todo}[1]{}
\newcommand{\argument}[1]{}
\fi

\usepackage{listings}

\lstdefinestyle{mystyle}{
    backgroundcolor=\color{backcolour},   
    commentstyle=\color{codegreen},
    keywordstyle=\color{magenta},
    numberstyle=\tiny\color{codegray},
    stringstyle=\color{codepurple},
    basicstyle=\ttfamily\footnotesize,
    breakatwhitespace=false,         
    breaklines=true,                 
    captionpos=b,                    
    keepspaces=true,                 
    numbers=left,                    
    numbersep=5pt,                  
    showspaces=false,                
    showstringspaces=false,
    showtabs=false,                  
    tabsize=2
}

\newtcolorbox[auto counter]{hypothesis}[1]
{
    coltitle=black,
    colframe=gray!25,
    arc = 0.5mm,
    title=\textbf{Hypothesis~\thetcbcounter:} (#1)
}

\AtBeginDocument{%
  \providecommand\BibTeX{{%
    \normalfont B\kern-0.5em{\scshape i\kern-0.25em b}\kern-0.8em\TeX}}}

\setcopyright{acmcopyright}
\copyrightyear{2020}
\acmYear{2018}
\acmDOI{ }

\acmConference[]{}{}{}
\acmBooktitle{}
\acmPrice{0.00}
\acmISBN{}

\begin{document}

\title{Applying system descriptors to address ambiguity on deployment diagrams}
\author{Jalves Nicacio}
\affiliation{%
  \institution{Université du Québec à Chicoutimi}	
  \city{Chicoutimi}
  \state{Québec}
  \country{Canada}
}
\email{jalves.mendonca-nicacio1@uqac.cam}

\author{Fabio Petrillo}
\affiliation{%
  \institution{Université du Québec à Chicoutimi}
  \city{Chicoutimi}
  \state{Québec}
  \country{Canada}
  }
\email{fabio@petrillo.com}

\renewcommand{\shortauthors}{Nicacio and Petrillo}

\begin{abstract}
  Communication between practitioners is essential for product quality in the DevOps context. This communication often takes place through deployment diagrams of a system under development. 
  However, it is common diagrams to become ambiguous or inconsistent as the system progresses and goes to a continuous delivery pipeline or production. Moreover, diagrams could not follow the evolution of systems, and it is challenging to associate diagrams to production. 
  In this paper, we propose the use of system descriptors to address the ambiguity of deployment diagrams. We state three main hypotheses (1) if a deployment diagram is generated from a valid system descriptor then the diagram is unambiguous; (2) if a valid system descriptor is generated from a deployment diagram then the descriptor is unambiguous; (3) if a diagram $\mu$ generated from a descriptor $A$ is unambiguous and if a descriptor $B$ is generated from the diagram $\mu$  equally unambiguous then descriptors $A$ and $B$ are equivalent.
  We report a case study to test our hypotheses. We constructed a system descriptor from Netflix deployment diagram, and we applied our tool to generate a new deployment diagram. Finally, we compare the original and generated diagrams to evaluate our proposal.
  Our case study shows the generated deployment diagrams are graphically equivalent to system descriptors and eliminated ambiguous aspects of the original diagram. Thus, our preliminary results lead to further evaluation in controlled and empirical experiments to test our hypotheses conclusively.
\end{abstract}



\maketitle

%
%
\section{Introduction}
\label{sec:introduction}

In a modelling process, several levels of abstraction can represent software systems across different resources, such as text or diagrams. Deployment diagrams are useful for communication and understanding of systems, as they motivate a more active discussion among participants as facilitating the memorization of details about systems \cite{Jolak2020}.

Nowadays, the growth and rapid adoption of DevOps are notorious. The benefits obtained by the software industry through continuous integration, testing, monitoring and delivery processes are undeniable. Additionally, the adoption of systems descriptors, such as Puppet, Chef, Docker-compose and Kubernetes Pods, proved to be a prominent approach to increase the quality and productivity of information systems.

However, system diagrams (as UML deployment diagrams) usually remain schematic and disassociated from production reality. Furthermore, diagrams do not follow the evolution of the systems, and there are challenges for engineers to synchronize the diagrams with the system in production.


Deployment diagrams are complex to design. Observing architecture diagrams from technical blogs of Amazon, Linkedin, or Netflix,  we noticed that engineers use general-purpose notations to create models in tools such as Visio or previous draw.io.  Furthermore, few deployment diagrams are created using UML (or partially using them). Brown \cite{Brown2020} states that it is common for engineers to reuse elements designed in other diagrams, using the copy-paste technique, but this can lead to making mistakes and inconsistencies. Moreover, the semantic and semiotic variation of graphical notations for the representation of systems architecture gives engineers the freedom to adapt graphical components to different domains, but it also allows ambiguous diagrams.

Thus, in this, paper, we propose the use of system descriptors to address the ambiguity of deployment diagrams. We state three main hypotheses (1) if a deployment diagram is generated from a valid system descriptor then the diagram is unambiguous; (2) if a valid system descriptor is generated from a deployment diagram then the descriptor is unambiguous; (3) if a diagram $\mu$ generated from a descriptor $A$ is unambiguous and if a descriptor $B$ is generated from the diagram $\mu$  equally unambiguous then descriptors $A$ and $B$ are equivalent.



To evaluate our hypotheses, we performed a case study from a single system descriptor: Docker-compose files. In this way, we have implemented a prototype capable of generating architecture diagrams from docker-compose files. The prototype reads all blocks of information and analyzes them according to a meta-model that transforms the tags in the docker-compose file into a visual element of the corresponding diagram.

The main contribution is to present our hypotheses and our preliminary evaluation, showing the generated deployment diagram could be equivalent to system descriptors and eliminate ambiguous aspects of the original diagram. The remainder of this paper is organized as follows:
Section \ref{sec:context} presents relevant concepts for this research (Section \ref{sec:concepts}) and defines and contextualizes the ambiguity on deployment diagram (Section \ref{sec:problem}).
Our proposal is formalized in Section \ref{sec:proposal}.
In Section \ref{sec:case-study}, we describe a study carried out with a tool prototype that generates deployment diagrams from Docker-Compose files.
In Section \ref{sec:preliminary-evaluation}, we present and discuss a preliminary evaluation.
Section \ref{sec:related-work} presents related work.
Section \ref{sec:conclusion} concludes our work and presents future works.

%
%
\section{Context}
\label{sec:context}

We divided this section into two subsections. Firstly, we present the most relevant concepts for this paper in \ref{sec:concepts}. Then, in Section \ref{sec:problem}, we introduce the definition of ambiguity on deployment diagram.

\subsection{Relevant concepts}
\label{sec:concepts}

\paragraph{\textbf{Systems descriptors.}} They are scripts for automation, standardization and management of infrastructure in production environments. 
System descriptors emerge within the context of Infrastructure as Code (IaC), which specifies the definition and set up of the software infrastructure required to run a system by using configuration scripts \cite{Artac2017}.

In practice, system descriptors are artifacts that describe a system architecture. Providers, such as Chef, Dockerfile and Puppet \cite{Chef2020, Docker2020, Puppet2020}, are tools that produce system descriptors. Likewise, container orchestrators, such as Docker-Compose and Kubernetes Pods \cite{Dockercompose2019, Kubernetes2020}, also produce system descriptors.


The system descriptor files are written in different languages, such as JSON, YAML, or even in a specific domain language (DSL), as is the case with the Puppet tool. They can still be written in general-purpose languages, as is the case with the Chef tool that uses Ruby to generate the system descriptor scripts.

The Listing \ref{lst:docker-compose-example} illustrates a system descriptor: a docker-compose script that details the configuration of the database service, called \textit{db}. According to the script's instructions, Docker-compose should build a docker application container, named \textit{mysql-container}, with the \textit{mysql} image available on dockerhub\footnote{\url{https://hub.docker.com/\_/mysql}} repository.

\begin{lstlisting}[caption={Docker-compose.yaml example}, label={lst:docker-compose-example}]
version: "3.7"
services:
  db:
    image: mysql
    command: --default-authentication-plugin=mysql_native_password
    container_name: mysql-container
    environment:
      MYSQL_ROOT_PASSWORD: secret
    volumes:
      - .api/db/data:/var/lib/mysql
    restart: always
\end{lstlisting}

While these languages are efficient in informing the computer of accurate information about the architecture and infrastructure of the system, diagrams do a better job when it comes to transmitting data to humans \cite{Flaatten2020}.

\paragraph{\textbf{Diagram as Code}} 
Diagram as Code is an approach to generate diagrams in the same way that IaC generates systems infrastructure: through programming.
According to \cite{Mingrammer2020},  Diagram as Code is the design of systems architecture diagrams, using a specific programming language to describe the diagram's elementsand their relationships.

This approach has sparked recent interest among software engineering practitioners. A non-exhaustive list of some web articles on the subject is given in \cite{Flaatten2020, Meyer2019, Mingrammer2020, Brown2020a}.
Table \ref{tab:diagram-as-code-tools} presents a list compiled by \cite{Flaatten2020} of Diagram as Code tools and that we extend by adding two more tools found on the Web: Diagrams \cite{Mingrammer2020}  and diagrams-as-code \cite{Meyer2019}.

\begin{table}[h]
\centering
\caption{"Diagram as code" tool list. The list has been extended from \cite{Flaatten2020}}
\label{tab:diagram-as-code-tools}
\resizebox{\columnwidth}{!}{%
\begin{tabular}{lllll}
\hline
Tool & Language & License & Local & Online \\ \hline
Graphviz & DOT & Eclipse Public License 1.0 & yes & yes \\
PlantUML & Text & GPL-3.0 & yes & yes \\
Mermaid & Text & MIT License & yes & yes \\
Ditaa & ASCII & LGPL-3.0 & yes & no \\
WSD & Text & - & no & yes \\
code2flow & Text & - & no & yes \\
Structurizr & Java, .NET & - & no & yes \\
Diagrams & Python & MIT License & yes & no \\
diagram-as-code & Javascript & - & yes & yes \\\bottomrule
\end{tabular}%
}
\end{table}

As far as we know, this is the first work to introduce the diagram as a code approach in the scientific community.

\paragraph{\textbf{Deployment diagrams}}
There is a vast literature on deployment diagrams. According to the UML Specification \cite{uml2017omg}, \enquote{deployment diagrams show the configuration of run-time processing elements and the software components, processes, and objects that execute on them}. Another definition says that a deployment diagram is a graph of nodes connected by communication associations \cite{chang2000}. One of the deployment diagram functions is to map the software architecture to the hardware \cite{Arlow2005}.

In this way, a deployment diagram is composed of nodes, communication associations and, where desired, dependency associations between nodes.  Living within the nodes are the run-time components and objects \cite{Swain2010}.

A node represents a physical resource, such as a computer, a router, or a printer. Nodes can contain other elements, such as components or artifacts \cite{ainapure2008}. Communication associations are connections stereotyped to show either a physical connection medium, such as fibre or software protocols (for example, TCP/IP or HTTP) .

\paragraph{\textbf{Ambiguity}} According to Cambridge dictionary, ambiguity is defined as the fact of something having more than one possible meaning and therefore possibly causing confusion\footnote{\url{https://dictionary.cambridge.org/pt/dicionario/ingles/ambiguity}}. 

Another word for ambiguity is amphibology, whose epistemological origin is the greek word \textit{amphibolos}\footnote{\url{https://www.merriam-webster.com/dictionary/amphibology}}. \textit{Amphibolos} is the combination of two other words: \textit{amphi}, which means "both" and \textit{ballein} which means "to throw". This word is literally translated from Greek as "encompassing" or "hitting at both ends", which is therefore a figuration for ambiguity.

In psychology, ambiguity in figures has been extensively investigated to reveal critical sensory, motor, cognitive and physiological processes involved in the perception of form. \cite{Long2004}.

\begin{figure}[ht]
    \centering
    \includegraphics[width=0.5\columnwidth]{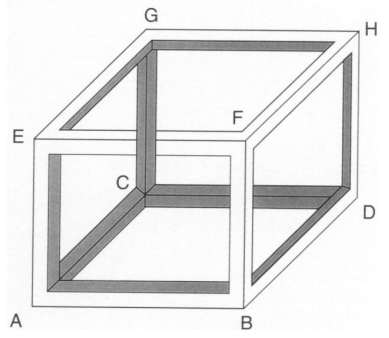}
    \caption{Impossible Cube of Escher \cite{Goldstein1996}}
    \label{fig:impossible-cube}
\end{figure}

Reversible figures and impossible figures are examples of images that explore the graphic similarity to lead the observer to the phenomenon of multi-stable perception or to the condition of logical ambiguity \cite{Long2004}. The Escher Cube, shown in Figure \ref{fig:impossible-cube} illustrates an example of an impossible figure \cite{Goldstein1996}.

\subsection{Ambiguity on deployment diagram}
\label{sec:problem}





As previously defined, ambiguity is the fact of something having more than one possible meaning. However, we are interested in the concept of ambiguity related to deployment diagrams.

On the best of our knowledge, there is not a clear definition or studies to evaluate ambiguity on deployment diagrams. Thus, we define \textbf{an deployment diagram is ambiguous when the graphical representation has more than one possible meaning}.

%

\begin{figure}[htp]
	\subcaptionbox{Original caption: \enquote{DBLog High Level Architecture}. From \cite{Andreakis2019a}}{%
		\includegraphics[clip,width=\columnwidth]{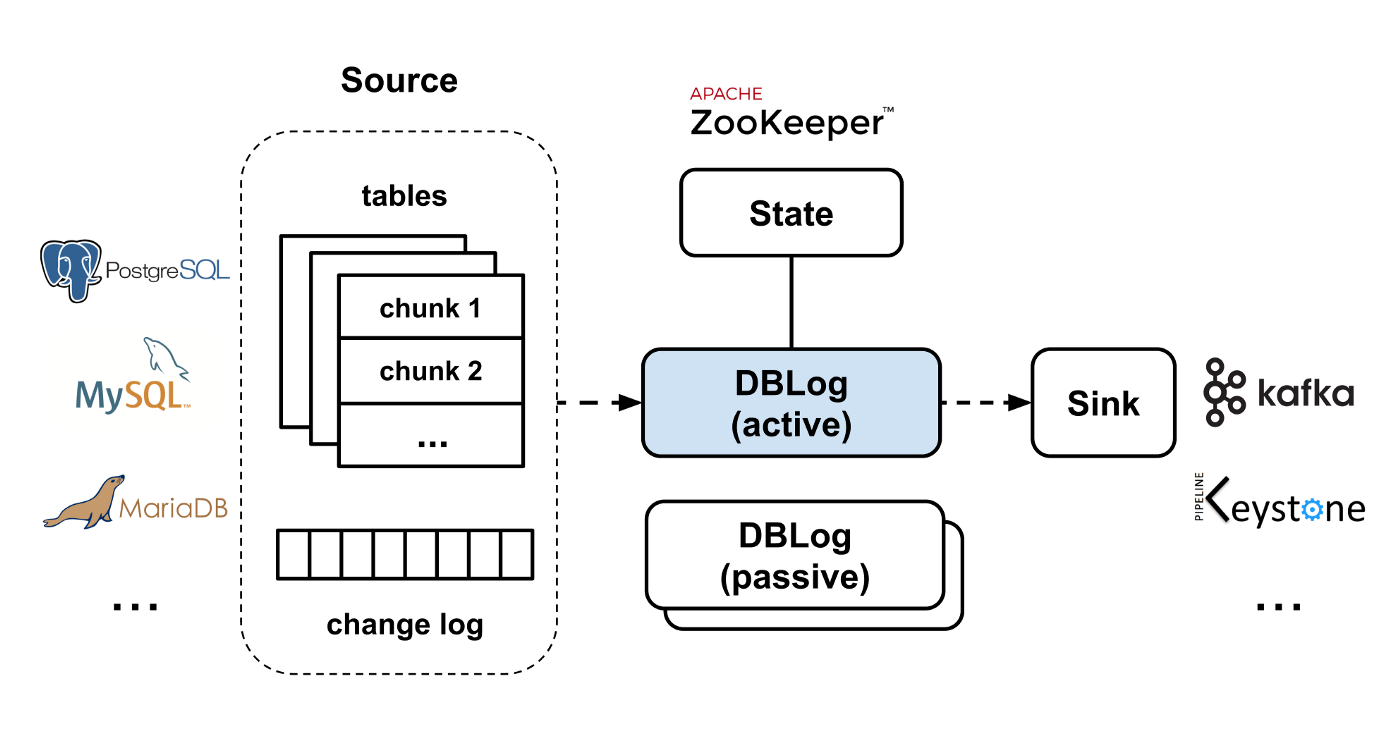}%
	}
	
	\subcaptionbox{Original caption: \enquote{Druid Cluster Overview}. From \cite{Sykes2020}}{%
		\includegraphics[clip,width=0.9\columnwidth]{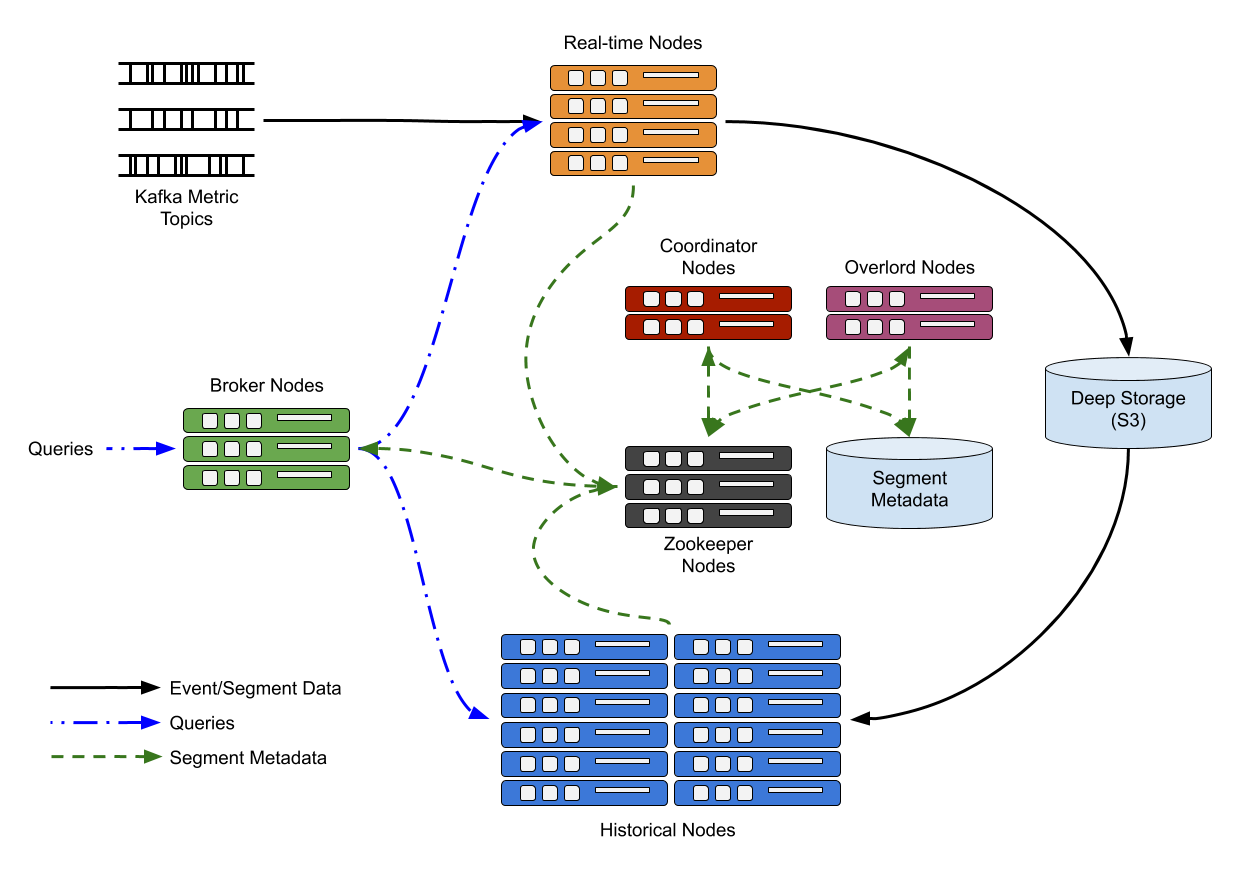}%
	}
	\caption{Examples of real-life system architecture diagram}
	\label{fig:diagrams-examples}
\end{figure}

To illustrate this problem, consider the diagrams in Figure \ref{fig:diagrams-examples}. Analyzing those diagrams, we could formulate several questions. For example, (1) how many services those systems have?; (2) where do those system are deployed?; (3) how much do they consistently communicate enough details to represent the same system using other languages or graphical notations, such as UML?

\begin{figure}[htp]
	\subcaptionbox{UML version of the diagram in Figure \ref{fig:diagrams-examples} (a)}{%
		\includegraphics[clip,width=\columnwidth]{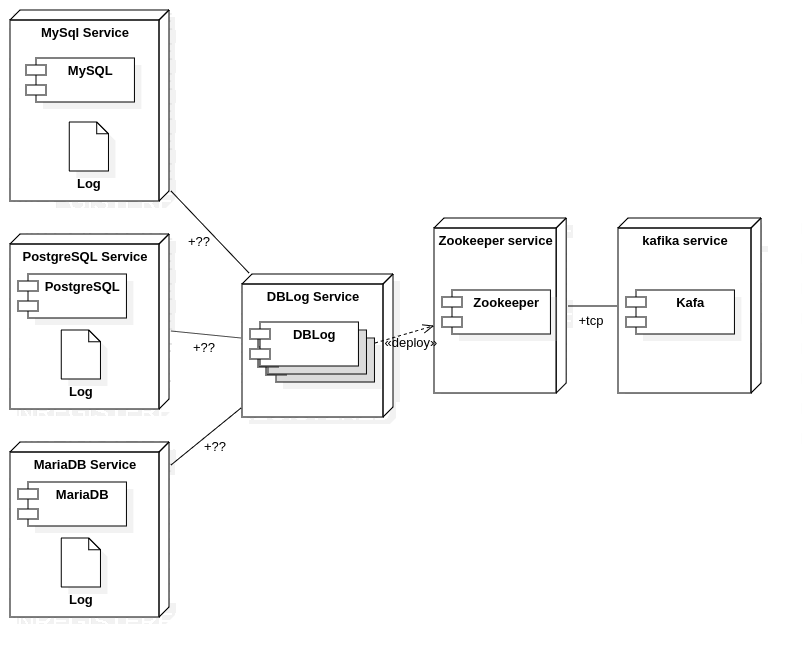}%
	}
	
	\subcaptionbox{UML version of the diagram in Figure \ref{fig:diagrams-examples} (b)}{%
		\includegraphics[clip,width=0.9\columnwidth]{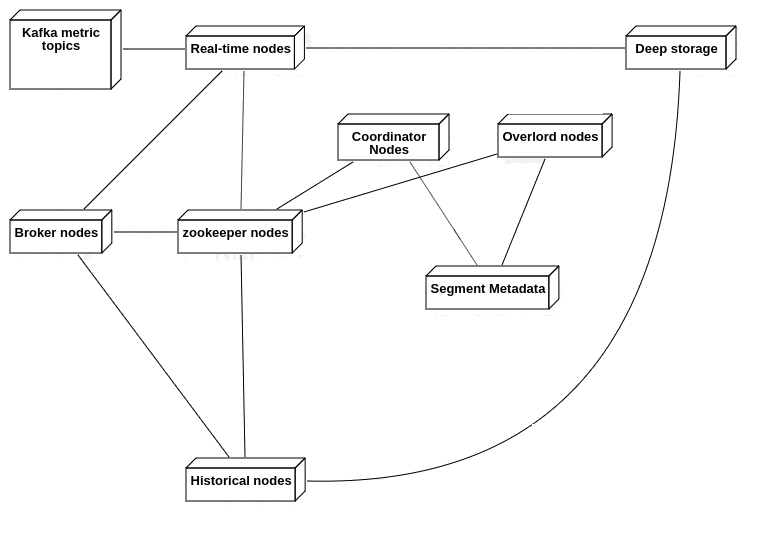}%
	}
	\caption{Examples of real-life system architecture diagram}
	\label{fig:diagrams-examples-uml}
\end{figure}

Figure \ref{fig:diagrams-examples-uml} presents the UML deployment diagrams that we tried to model from the original diagrams. Firstly, we noticed some problems related to understanding. For example, the diagrams are context-dependent. It is practically impossible to analyze and understand the diagrams without a detailed reading of the text in which the diagram is inserted.
However, one of the most significant issues observed was that the semiotics used in the diagrams is ambiguous or not very representative. 

%

We investigated dozens of deployment diagrams of industry systems\footnote{Our future work is to investigate systematically our large data set of deployment diagrams from the grey literature.}, all extracted from technical blogs and company websites, such as Amazon, Linkedin, Netflix, Spotify, and others. For each diagram analyzed, at least one of these issues occurs, making the diagram ambiguous:

\paragraph{\textbf{Issue 1}: Use of boxes associated with colours and different border styles.}
Rectangles or squares are generally associated with colours to separate or classify different types of elements. Also, many designers differ boxes by varying the border style (solid, dotted, or dashed). These specifications are not always provided in a caption, which makes understanding difficult.

\begin{figure}[ht]
    \centering
    \begin{subfigure}{0.25\textwidth}
        \includegraphics[width=0.7\linewidth]{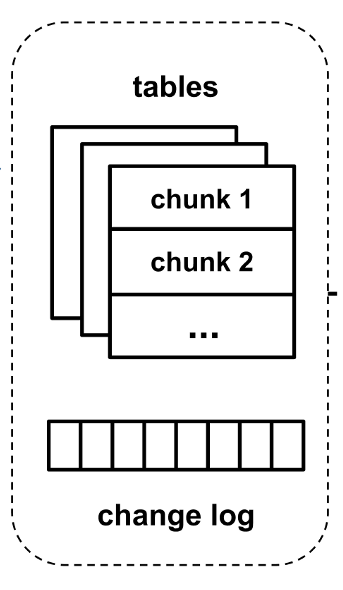} 
        \caption{Dashed region}
        \label{fig:region}
    \end{subfigure}
    \begin{subfigure}{0.2\textwidth}
        \includegraphics[width=0.9\linewidth]{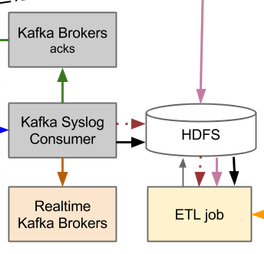}
        \caption{Squares of different colours}
        \label{fig:squares_colors}
    \end{subfigure}
    
    \caption{Use of squares and rectangles: use of colours without legend}
    \label{fig:symbol_squares}
\end{figure}

As noted in Figure \ref{fig:symbol_squares}, it is difficult to understand how each quadrilateral is classified, raising questions such as: \textit{``Why this is grey, and that is yellow?"} (Fig \ref{fig:squares_colors}), or \textit{``Boxes with dashed borders represent a logical group of components or a closer view (zoom in) of a component?"} (Figure \ref{fig:region});

\paragraph{\textbf{Issue 2}: Use of arrows.}
It is not always clear what the arrow represents in the diagram. For example, depending on what the engineer suggested, the arrows can represent a flow of data or connections between components. Arrows are also used in conjunction with different colours and shapes (dashed, dotted or continuous lines).

\begin{figure}[ht]
    \centering
    \includegraphics[scale=0.5]{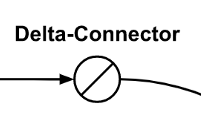}
    \caption{Failure in semiotic representation: the cut circle is generally used to represent prohibition}
    \label{fig:delta_connector}
\end{figure}

\paragraph{\textbf{Issue 3}: Failure in semiotic representation}
In various symbol systems, the cut circle is generally used to represent prohibition. In the diagram in Figure \ref{fig:delta_connector} the cut circle represents a connection interface.

%
%
\section{The proposal}
\label{sec:proposal}

In view of all the arguments already presented and the problem addressed in section \ref{sec:problem} and to produce clear diagrams, we propose the use of system descriptors to address the ambiguity of deployment diagrams, generating and validating them. Thus, we formulate the following hypotheses:



\begin{hypothesis}{the deployment diagram unambiguity}
    If a deployment diagram is generated from a valid system descriptor, then the diagram is unambiguous.
    \label{sec:h1}
\end{hypothesis}

We assume that a descriptor file is naturally unambiguous for two reasons: (1) System descriptor files are written in a formal language, such as YAML (used by Docker-compose and Kubernetes Pods) or Ruby (used by Chef); (2) be processed automatically by a finite-state machine. 
If the diagram generated from system descriptors presents each item of the descriptor, this diagram will also be unambiguous.




\begin{hypothesis}{the system descriptor unambiguity}
    If a valid system descriptor is generated from a deployment diagram, then the descriptor is unambiguous.
    \label{sec:h3}
\end{hypothesis}

We state that if a deployment diagram has accurate enough data to generate a system descriptor that is valid and executable in a tool that processes system descriptor files, then the generated descriptor is unambiguous. Transitively, the diagram is unambiguous either. 


\begin{hypothesis}{the equivalency of descriptors}
    If a diagram $\mu$ generated from a descriptor $A$ is unambiguous (\textit{Hypothesis 1}) and if a descriptor $B$ is generated from the diagram $\mu$  equally unambiguous (\textit{Hypothesis 2}) then descriptors $A$ and $B$ are equivalent.
    \label{sec:h3}
\end{hypothesis}



%
%
\section{Case Study}
\label{sec:case-study}

We conducted a case study focused on testing our hypotheses. We divide this case study into two steps.

\begin{figure}[h]
    \centering
    \includegraphics[width=\columnwidth]{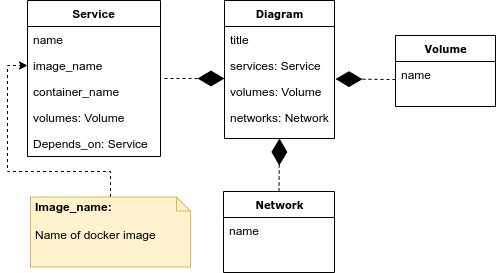}
    \caption{The Docker-compose transformation meta-model}
    \label{fig:compose-model}
\end{figure}

In the \textbf{first step}, we use the GitHub Awesome-compose\footnote{\url{https://github.com/docker/awesome-compose}} repository to create a model to describe \textbf{docker-compose} system descriptors. We selected ten docker-compose files and generated deployment diagrams using the \textit{Diagram as Code}\footnote{\url{https://diagrams.mingrammer.com/}} tool. Thus, from the analysis of the ten generated diagrams, we developed a model to generate deployment diagrams from docker-compose files. Figure \ref{fig:compose-model} shows our docker-compose model. 

The model corresponds to docker-compose specification 3.8 for Docker engine release	19.03.0 \cite{Dockercompose2019}, which defines three top-level entities: services, volumes and networks. The service entity contains the settings that are applied to each container within a service.
A volume describes the permanent storage for access in a service, and Network describes the logical networking in a container.


We developed a tool that generates deployment diagrams named
\textbf{Descriptor to Deployment Diagram} or \textbf{D2DD}\footnote{\url{https://github.com/jalvesnicacio/descriptors-diagrams}}.
This tool has a system descriptor file as input, and a deployment diagram and a Diagram as Code script as output.

In its current version, D2DD only recognizes docker-compose as a system descriptor and uses the Diagrams\footnote{\url{https://diagrams.mingrammer.com/}} library to generate deployment diagrams in the Diagram as Code style. 

\begin{figure}[ht]
    \centering
    \includegraphics[width=0.9\columnwidth]{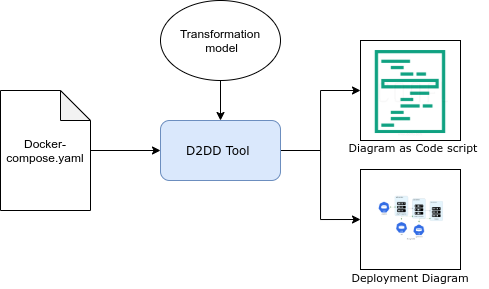}
    \caption{D2DD tool overview}
    \label{fig:prototype}
\end{figure}

Figure \ref{fig:prototype} presents an overview of how the D2DD tool works. The tool reads all blocks of information in docker-compose input file, analyzes them according to the transformation meta-model (see Sec. \ref{sec:case-study}) and transforms each tag in the input file into visual elements of the corresponding deployment diagram. D2DD tool also generates the equivalent diagram as code script simultaneously.

\begin{figure}[hb]
    \centering
    \includegraphics[width=\columnwidth]{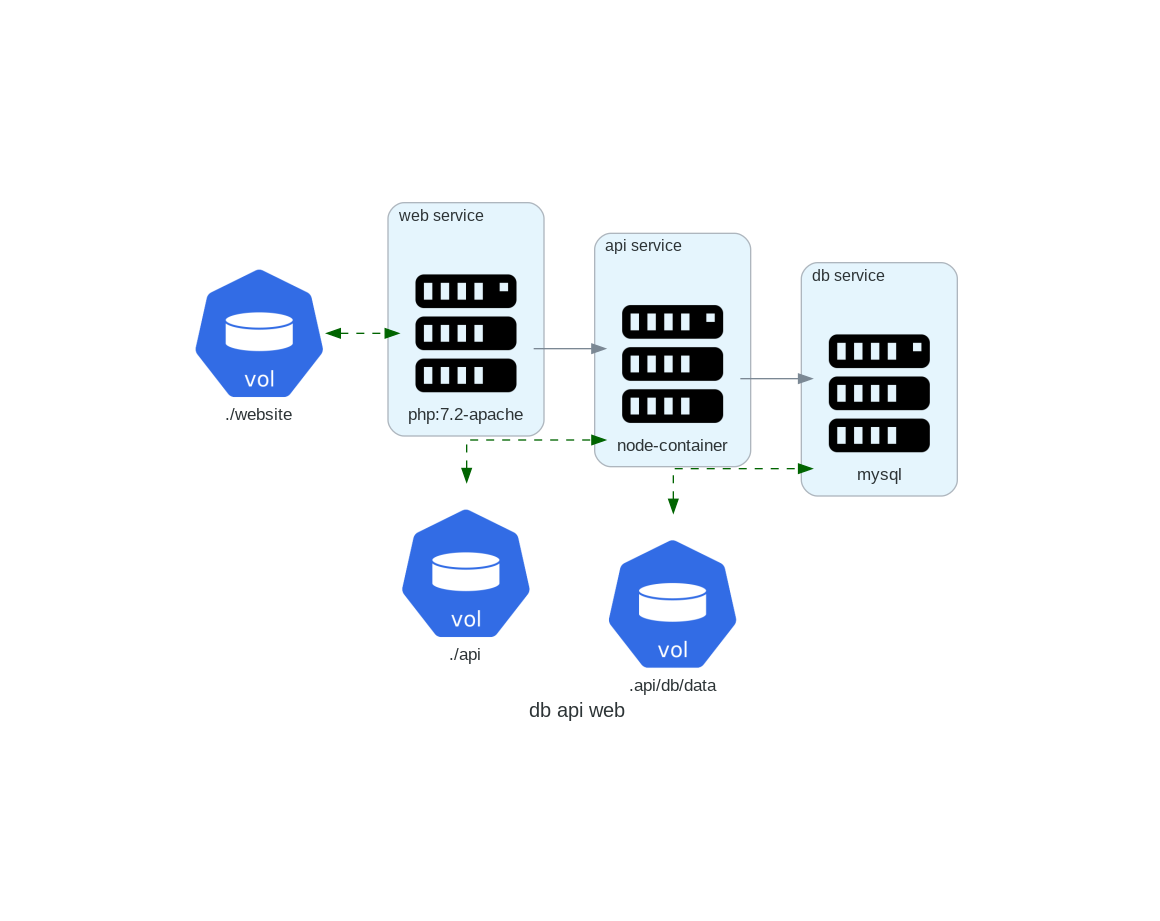}
    \caption{Diagram generated by the prototype}
    \label{fig:generated_diagram}
\end{figure}

An example of a deployment diagram generated by the tool is illustrated in Figure \ref{fig:generated_diagram}. This deployment diagram  details the configuration of tree services, Apache+PHP, NodeJs and MySQL. It informs which docker image must be instantiated in each service.

In the \textbf{second step} we are interested in comparing the deployment diagrams for real-world systems with deployment diagrams for those same systems, but this time, generated through system descriptors.
To achieve our goal, we created a docker-compose.yaml file (Figure \ref{fig:diagrams-examples} (a)), and we used D2DD to automatically generate deployment diagram from the docker-compose file by using our transformation meta-model (Figure \ref{fig:compose-model}).



The descriptor represent the closest representation of the system from the blog description and the figure.
Listing \ref{lst:docker-compose} presents the file descriptor for the DBLog system \cite{Andreakis2019a} whose architecture is represented in Figure \ref{fig:diagrams-examples}(a).

\begin{lstlisting}[caption={Docker-compose.yaml example}, label={lst:docker-compose}]
version: '3.8'
services:
  mysql:
    image: mysql
    ports:
     - 3306:3306
    environment:
     - MYSQL_ROOT_PASSWORD=secret
     - MYSQL_USER=mysqluser
     - MYSQL_PASSWORD=mysqlpw
    volumes:
      - db-data:/var/lib/mysql
  postgres:
    image: postgres:9.4
    volumes:
      - db-data:/var/lib/postgresql/data
  dblog:
    build:
      context: api
      dockerfile: Dockerfile
      container_name: dblog
    restart: always
    ports:
      - "9001:9001"
    depends_on:
      - mysql
      - postgres
    links:
      - zookeeper
  zookeeper:
    image: debezium/zookeeper:${DEBEZIUM_VERSION}
    ports:
     - 2181:2181
     - 2888:2888
     - 3888:3888
  kafka:
    build:
      context: kafka
      dockerfile: Dockerfile
      container_name: kafka
    links:
      - zookeeper
    ports:
      - "9092:9092"
    environment:
      KAFKA_ADVERTISED_HOST_NAME: $CF_HOST_IP
      KAFKA_ZOOKEEPER_CONNECT: zk:2181
      KAFKA_MESSAGE_MAX_BYTES: 2000000
      KAFKA_CREATE_TOPICS: "Topic1:1:1"
    volumes:
      - /var/run/docker.sock:/var/run/docker.sock
    depends_on:
      - zookeeper
volumes:
  db-data:
\end{lstlisting}

Once the system descriptor presented in Listing  \ref{lst:docker-compose} is provided as input, the developed tool delivers both a deployment diagram and a Diagram as Code script. Listing \ref{lst:diagram-as-code-generated} shows the Diagram as Code script automatically generated by the tool  when it receives the script from Listing \ref{lst:docker-compose} as input. 

\begin{figure}[h]
    \centering
    \includegraphics[width=\columnwidth]{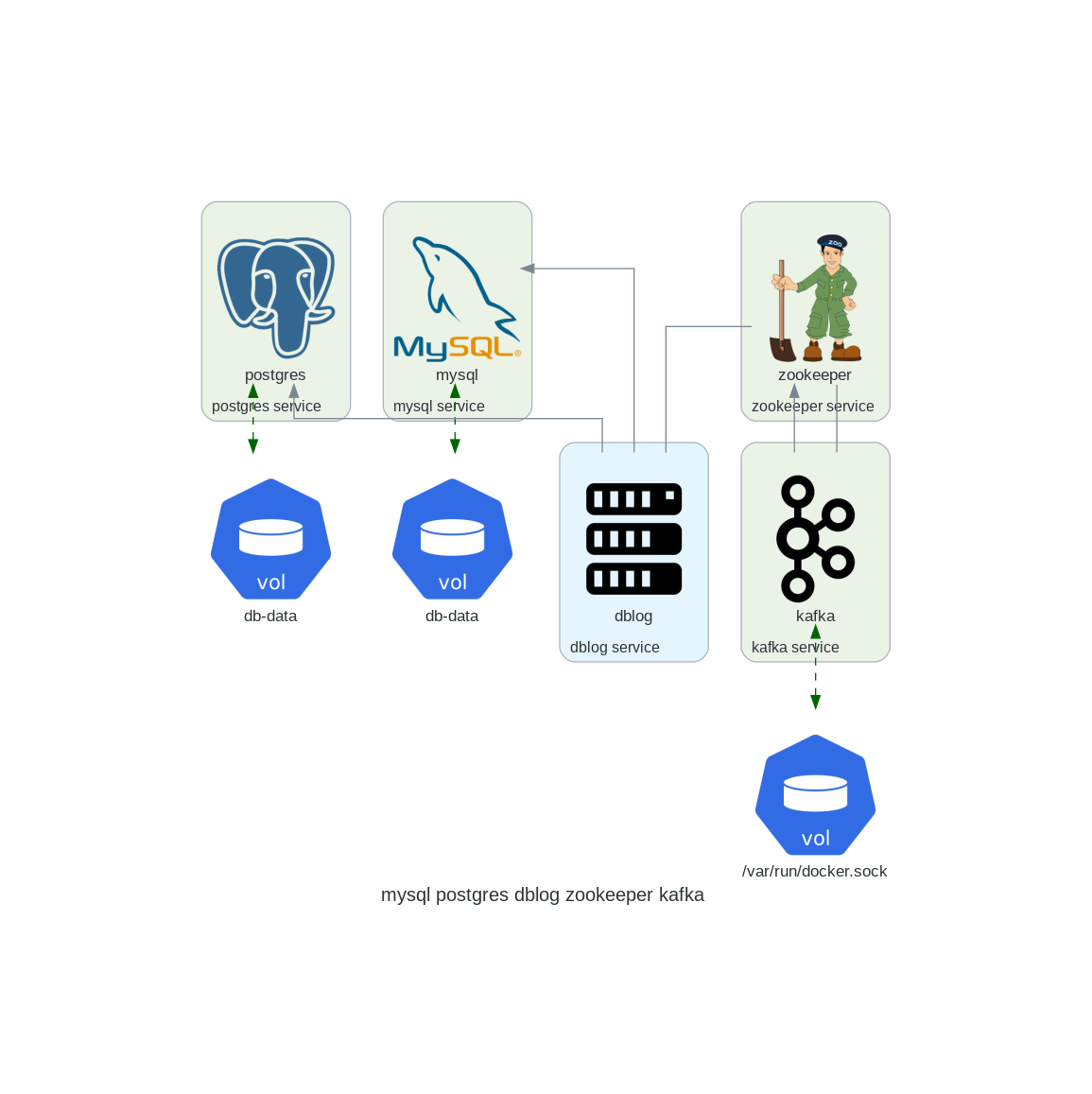}
    \caption{Deployment diagram corresponding to image \ref{fig:diagrams-examples} (a), generated from the D2DD tool and modified manually to add some semantic aspects.}
    \label{fig:dac-diagram-neflix}
\end{figure}

Figure \ref{fig:dac-diagram-neflix} shows deployment diagram corresponding to that of Figure \ref{fig:diagrams-examples}(a). The diagram was manually adapted from the diagram generated by the tool to include semantic aspects. We simply modify the order in which the elements appear in the diagram so that elements that have similar functions remain together, such as database systems. In addition, to enrich the deployment diagram, we include the logos of the known software (MySQL, PostgreSQL, Zookeeper and Kafka).

\begin{lstlisting}[caption={Script Diagram as Code generated by the D2DD tool}, label={lst:diagram-as-code-generated},language=Python]
from diagrams import Cluster, Diagram as DaC, Edge
from service import Service
from diagrams.k8s.storage import Volume
from diagrams.onprem.compute import Server

with DaC("mysql postgres dblog zookeeper kafka ", filename= "./diagram-adhoc", show=False, direction="TB"):
    with Cluster("mysql service"):
        mysql = Server("mysql")
    with Cluster("postgres service"):
        postgres = Server("postgres")
    with Cluster("dblog service"):
        connect = Server("connect")
    with Cluster("zookeeper service"):
        zookeeper = Server("zookeeper")
    with Cluster("kafka service"):
        kafka = Server("kafka")
    kafka >> zookeeper
    connect - zookeeper
    connect - mysql
    connect - postgres
    kafka - zookeeper
    vol_mysql = Volume("db-data")
    vol_mysql >> Edge(color="darkgreen", style="dashed") << mysql
    vol_postgres = Volume("db-data")
    vol_postgres >> Edge(color="darkgreen", style="dashed") << postgres
    vol_kafka = Volume("/var/run/docker.sock")
    vol_kafka >> Edge(color="darkgreen", style="dashed") << kafka
\end{lstlisting}

\section{Preliminary evaluation}
\label{sec:preliminary-evaluation}

We compared the deployment diagrams extracted from technical blogs with deployment diagrams generated using D2DD (Figure \ref{fig:diagrams-examples} \textit{versus} Figure \ref{fig:dac-diagram-neflix}). From that analysis, we made the following observations:

\begin{tcolorbox}[colframe=gray!50, coltitle=black, title=Observation 1]
    Generated deployment diagrams eliminate same ambiguous aspects of the original diagram.
\end{tcolorbox}

We observed that some elements of the original deployment diagram in Figure \ref{fig:diagrams-examples}(a) do not appear in the diagram generated in the case study, such as "State" and "Sink" boxes.  Even understanding that these elements are part of the fundamental concepts of the Zookeeper and Kafka tools, when inserted in the deployment diagram, they cause misunderstanding, since their graphical notations are very similar to the graphical notation of the DBLog boxes, which represents the main application of the system.

Another element that does not appear in the generated deployment diagram is the dashed region around the tables and the database's change logs. If we isolate the deployment diagram from the blog text, this region could provide different interpretations, such as a logical connection between tables and change logs, an expanded view (zoom magnification) of the databases or a semantic organization stating that the elements within the dashed region have similar functions.

As we observe in Figure \ref{fig:dac-diagram-neflix}, the use of a defined system descriptor (docker-compose in that case) to generate the diagram using D2DD eliminated all elements that carry more than one meaning. However, we note that the topological organization of the elements (that is, the order in which the elements appear in the diagram) is important for understanding the diagram because it expresses a semantic organization of the elements.


\begin{tcolorbox}[colframe=gray!50, coltitle=black, title=Observation 2]
    Generated deployment diagram could be equivalent to system descriptors
\end{tcolorbox}

We observed that, in Figure 6, the direction of the arrows has been inverted concerning the arrows in the diagram in Figure 1. In graph theory, the orientation of the arrows refers to a dependency relationship between the two connected elements (asymmetric digraphs concept). In this way, we say, for example, that the relationship between the DBLog service and the MySQL service is expressed as \textit{"The DBLog service depends on the MySQL service"}, which is exactly what the system docker-compose file tells us (see line 26 of Listing \ref{lst:docker-compose}).

Finally, we note that all docker-compose elements, such as services, volumes, dependency relationships or links, are unequivocally represented by a uniform and consistent graphical notation in the deployment diagram generated by the D2DD tool.

\section{Related Work}
\label{sec:related-work}

To the best of our knowledge, this is the first work that shows an approach to reduce ambiguity in deployment diagrams using system descriptors. Possibly, the closest studies are from Paraiso \textit{et al.} \cite{Paraiso2016} and Burco \textit{et al.} \cite{Burco2020}. The first study proposes an approach to modelling Docker containers in a way that guarantees its deployability and management. The second study proposes a formal model for evaluating and verifying the properties of container-based systems using Bigraphical Reactive Systems.

Concerning ambiguity in diagrams, more three articles were found that, to a certain degree, are correlated with our work \cite{Satish2010, Abbas2019, Piao2006}. However, they proposed approaches to validate diagrams based on the use of representation models. Differently, what we proposed an approach that applies a valid descriptor to generate deployment diagram.

%
%
\section{Conclusion}
\label{sec:conclusion}

This paper presents our proposal to use of system descriptors to address the ambiguity of deployment diagrams. We discussed in the Section \ref{sec:problem} about the ambiguity of diagrams, and we observe that the \textit{ad-hoc} systems modelling with the current diagramming tools are not able of address ambiguity in deployment diagrams.

Hence, we state three main hypotheses formulated from our observations in real-world deployment diagrams. Also, we present D2DD, a tool that uses Diagram as Code to design diagrams from Docker-Compose files.

We performed a case study to evaluate our hypothesis. Our case study shows generated deployment diagrams are graphically equivalent to system descriptors, and it does not preserved ambiguous aspects of the original diagram. Such observations lead to further evaluation in controlled and empirical experiments to test our hypotheses conclusively.



As future work, we will investigate the development of meta descriptors capable of generating platform-specific descriptors.


\balance
\bibliographystyle{ACM-Reference-Format}
\bibliography{main}
\end{document}